\documentclass[sigconf]{acmart}
\usepackage[utf8]{inputenc}
\usepackage{newunicodechar}
\newunicodechar{～}{\textasciitilde}
\renewcommand\footnotetextcopyrightpermission[1]{} 
\usepackage[font=small,labelfont=bf]{caption}
\usepackage{multirow}
\usepackage{enumitem}
\usepackage{subcaption}
\usepackage{graphicx}
\usepackage[skip=4pt]{caption} 
\usepackage{subcaption}
\usepackage{bbm}
\usepackage{amsmath}      

\usepackage{amssymb}      
\usepackage{algorithm}    
\usepackage{algpseudocode}
\usepackage{graphicx}     
\usepackage{enumitem}     
\usepackage{makecell}

\graphicspath{{Figures/}}

\newcommand{\ModelName}{$\mathrm{GMFlowRec}$}

\AtBeginDocument{%
  }

\setcopyright{acmlicensed}
\copyrightyear{2018}
\acmYear{2018}
\acmDOI{XXXXXXX.XXXXXXX}

\acmISBN{978-1-4503-XXXX-X/18/06}

\begin{document}

\title{Gaussian Mixture Flow Matching with Domain Alignment for Multi-Domain Sequential Recommendation}

\author{Xiaoxin Ye}
\authornote{Corresponding author.}
\affiliation{%
  \institution{University of New South Wales}
  \city{Sydney}
  \country{Australia}
}
\email{xiaoxin.ye@unsw.edu.au}

\author{Chengkai Huang}
\authornotemark[1] 
\affiliation{%
  \institution{University of New South Wales \& Macquarie University}
  \city{Sydney}
  \country{Australia}
}
\email{chengkai.huang1@unsw.edu.au}

\author{Hongtao Huang}
\affiliation{%
  \institution{University of New South Wales}
  \city{Sydney}
  \country{Australia}
}
\email{hongtao.huang@unsw.edu.au}

\author{Lina Yao}
\affiliation{%
  \institution{University of New South Wales \& CSIRO's Data61}
  \city{Sydney}
  \country{Australia}
}
\email{lina.yao@unsw.edu.au}


\begin{abstract}
Users increasingly interact with content across multiple domains, resulting in sequential behaviors marked by frequent and complex transitions. While Cross-Domain Sequential Recommendation (CDSR) models two-domain interactions, Multi-Domain Sequential Recommendation (MDSR) introduces significantly more domain transitions, compounded by challenges such as domain heterogeneity and imbalance. Existing approaches often overlook the intricacies of domain transitions, tend to overfit to dense domains while underfitting sparse ones, and struggle to scale effectively as the number of domains increases. We propose \textit{GMFlowRec}, an efficient generative framework for MDSR that models domain-aware transition trajectories via Gaussian Mixture Flow Matching. GMFlowRec integrates: (1) a unified dual-masked Transformer to disentangle domain-invariant and domain-specific intents, (2) a Gaussian Mixture flow field to capture diverse behavioral patterns, and (3) a domain-aligned prior to support frequent and sparse transitions. Extensive experiments on JD and Amazon datasets demonstrate that GMFlowRec achieves state-of-the-art performance with up to 44\% improvement in NDCG@5, while maintaining high efficiency via a single unified backbone, making it scalable for real-world multi-domain sequential recommendation.

\end{abstract}




%

\maketitle

\begin{figure}
    \centering
    \includegraphics[width=0.97\linewidth]{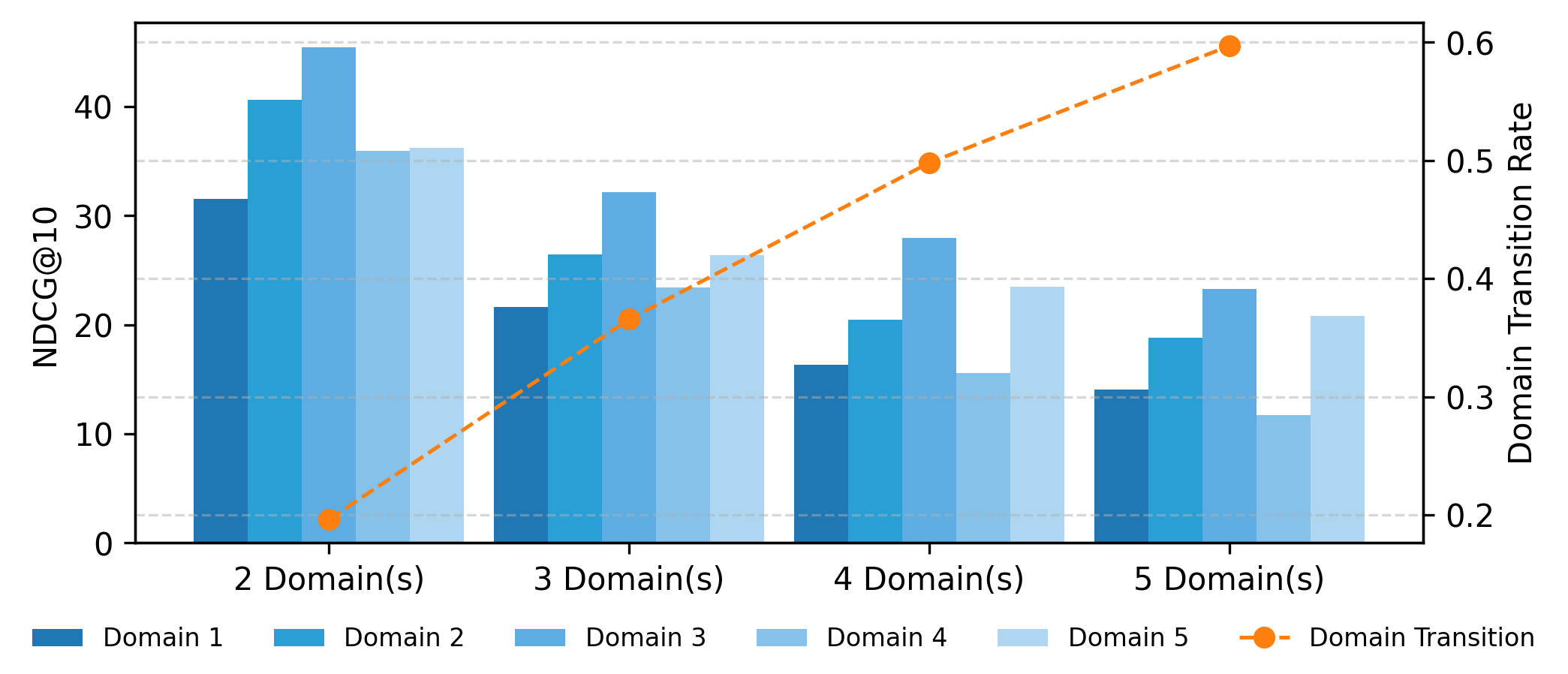}
    \caption{Performance of SASRec\cite{SASrec} on the JD multi-domain dataset. Bars report NDCG@10 across domains grouped by the number of domains within a user sequence. The line chart represents the domain transition rate, illustrating the increasing complexity of cross-domain interactions as the number of domains increases.}
    \label{fig:DomainTransition}
    \vspace{-1em}
\end{figure}

\section{Introduction}

On modern e-commerce platforms (e.g., Amazon, Alibaba, JD), users often browse across multiple domains (e.g., books, electronics, and movies) \cite{huang2024foundation,huang2025survey,huang2025towards,huang2025pluralistic,pan2025counterfactual,zhou2023contrastive}. These cross-domain transitions, shaped by dynamic user intents, give rise to behavioral sequences that are sparse, heterogeneous, and highly nonlinear, posing significant challenges to modeling user preferences in multi-domain recommendation \cite{cdsrsurvey}.

Traditional sequential recommenders typically operate within a single domain, capturing domain-specific temporal dependencies~\cite{huang2023dual,huang2023modeling}. Cross-Domain Sequential Recommendation (CDSR) methods extend this by modeling transitions between two domains, enabling limited cross-domain knowledge transfer. However, real-world platforms involve interactions across \emph{multiple} domains, demanding a more generalized paradigm, \textbf{Multi-Domain Sequential Recommendation (MDSR)}, where user behavior spans multiple domains and frequent transitions reflect complex intent evolution (e.g., browsing electronics before purchasing apparel). Scaling from CDSR to MDSR is far from trivial, introducing a profound \emph{scalability bottleneck} and new forms of data sparsity.

Unlike CDSR, which considers only pairwise domain transitions, MDSR must capture holistic user preferences and a combinatorially richer space of \textbf{domain transitions}, shifts in user activity from one domain to another (e.g., moving from browsing electronics to purchasing apparel). Such transitions are pervasive in large-scale platforms like e-commerce marketplaces or streaming services, where users fluidly switch across diverse categories.  

Our empirical study on the JD multi-domain dataset shows a sharp drop in NDCG@10 as domain transition rates increase—from 20\% in two-domain sequences to over 60\% in five-domain ones (Figure~\ref{fig:DomainTransition}). This trend underscores two key challenges:


\textbf{Challenge 1: Exponential Domain Transitions.} Expanding to multi-domain settings dramatically increases the number of possible transition paths. While CDSR involves only two bidirectional transitions (e.g., $A\!\rightarrow\!B$, $B\!\rightarrow\!A$), MDSR must handle a transition space that grows quadratically with the number of domains. This explosion renders most transitions infrequent and unpredictable, introducing a new sparsity form we call \emph{Transition Cold-Start}, where shifts between rarely paired domains lack sufficient training examples. Existing CDSR methods (e.g., AutoCDSR~\cite{AutoCDSR}) incorporate transition cues into attention mechanisms but fail to route knowledge effectively across this expanded transition space. This necessitates models capable of dynamically adapting to limited target-domain and transition-specific data \cite{liu2021improved,liu2021semi,liu2024boosting}.

\textbf{Challenge 2: Domain Heterogeneity and Imbalance}. Users interact with semantically distinct domains (e.g., health products for wellness, sports equipment for fitness, books for leisure) that vary in both purpose and interaction density. Modeling such heterogeneous behaviors with a single unified embedding, as in AutoCDSR~\cite{AutoCDSR}, risks conflating unrelated signals. Prior methods (e.g., DREAM~\cite{DREAM}, MDSRDSL~\cite{MDSRDSL}) either overlook multi-interest dynamics and rely on multiple encoders, which scale poorly and underfit sparse domains. Moreover, MDSR suffers from \emph{domain imbalance}, where many dense and sparse domains coexist, resulting in \textbf{few-shot sequences} (e.g., frequent health-product interactions but few in sports). Addressing these challenges jointly and efficiently remains an open problem. 

Diffusion models have recently demonstrated strong potential in capturing complex user preference distributions. However, a critical gap remains in multi-domain sequential recommendation (MDSR). Existing approaches such as DiffuRec~\cite{diff4serec} and FMRec~\cite{fmrec} are confined to single-domain intents, while cross-domain variants (e.g., DMCDR, CDCDR) often neglect sequential dependencies. Moreover, these models typically adopt a uni-modal Gaussian prior, implicitly assuming user preferences follow a single continuous mode. This mirrors the limitations of unified embeddings in multi-interest models~\cite{remit2023}, where diverse latent intents are collapsed into a single representation, restricting expressiveness and hindering generalization. As a result, uni-modal priors struggle to capture the coexisting, domain-sensitive interests in MDSR.

A further challenge arises from the training dynamics of diffusion-based recommenders. These models rely on slow, iterative denoising steps, which limit scalability. ADRec~\cite{ADRec} highlights a critical issue in this context: embedding collapse during end-to-end training. In recommendation, models initialize item embeddings randomly, with limited inherent meaning, making them vulnerable to collapse, severely degrades performance. Flow matching offers a promising alternative by learning deterministic mappings and supporting flexible prior distributions. However, without domain-aware conditioning, flow-based models still fall short in capturing multi-modal, domain-sensitive transitions, a core challenge in MDSR.

To bridge these gaps, we propose \textbf{\ModelName{}}, a domain-aware generative framework for MDSR. \ModelName{} integrates a \textbf{dual-masking encoder} that disentangles domain-invariant and domain-specific intents within a unified backbone, yielding domain-aware priors. These prior guide a \textbf{Gaussian Mixture Flow Matching} process to model multi-modal transition trajectories (e.g., books $\rightarrow$ electronics vs. books $\rightarrow$ movies), enabling scalable, transition-resilient recommendation. \textbf{Our key contributions are:}
\begin{itemize}[leftmargin=*]
    \item We present \textbf{\ModelName{}}, to our best knowledge, which is the first generative MDSR framework unifying Gaussian Mixture Flow Matching with domain-aware representation learning.
    \item We design a dual-masking encoder that disentangles domain-invariant and domain-specific intents in a shared backbone, providing efficient and adaptive generative guidance.
    \item Extensive experiments on large-scale Amazon and JD datasets show that \ModelName{} consistently outperforms state-of-the-art baselines, particularly under high-transition and few-shot domain sequence conditions, while maintaining efficiency.
\end{itemize}

\section{Related Work}

We review related literature in three areas: Cross-Domain Sequential Recommendation (CDSR), Multi-Domain Sequential Recommendation (MDSR), and recent Generative Models in Recommendation.

\textbf{Cross-Domain Sequential Recommendation.} CDSR aims to improve the accuracy of the recommendation by transferring patterns of user behavior between two domains while mitigating negative transfer. Early works like DA-GCN~\cite{DAGCN}, PSJNet~\cite{PSJNet} and $\pi$-Net~\cite{pinet} used gating mechanisms for selective knowledge transfer. Graph-based models such as C$^2$DSR~\cite{C2DSR}, and C2DREIF~\cite{C2DREIF} captured inter-domain dependencies via GNNs. Transformer-based approaches, including RecGURU~\cite{RecGURU} and DREAM~\cite{DREAM}, directly modeled sequential dynamics across domains, while ABXI~\cite{ABXI} introduced low-rank domain-invariant representations. However, these methods do not scale naturally to multiple domains.

\textbf{Multi-Domain Sequential Recommendation.} MDSR extends CDSR to three or more domains, introducing challenges such as increased domain heterogeneity, negative transfer, and scalability. AutoCDSR~\cite{AutoCDSR} employs automated transfer control to suppress harmful influence, while MDSR-DSL~\cite{MDSRDSL} embeds domains into a latent space to capture inter-domain similarity. CGRec~\cite{CGRec} frames knowledge sharing as a cooperative game, and SyNCRec\cite{SyNCRec} coordinate single- and cross-domain models to preserve domain specialization. However, most MDSR models focus on reducing negative transfer and lack mechanisms to capture multi-modal user preferences. Moreover, their reliance on multiple domain-specific transformers raises scalability concerns in large-scale applications.

\textbf{Generative Models in Recommendation.} Generative models such as diffusion~\citep{DhariwalN21, RombachBLEO22} and flow-based approaches~\citep{flowGen, schusterbauer2025diff2flow} have recently gained traction for their expressive modeling capabilities by reversing noise processes~\citep{YangZSHXZZCY24}. In recommendation, diffusion-based methods like DiffuRec~\citep{diff4serec}, DCRec~\citep{huang2024dual}, and DreamRec~\citep{DreamRec} demonstrate strong performance in sequential tasks. Extensions to cross-domain recommendation, such as DMCDR~\citep{DMCDR} and CDCDR~\citep{CDCDR}, incorporate domain conditioning but largely neglect sequential dynamics. Moreover, most of these models rely on uni-modal Gaussian assumptions, which constrain their ability to capture diverse user intents and transitions. Diffusion models also suffer from slow sampling and scalability limitations, making them less practical for large-scale multi-domain scenarios.

Flow matching~\citep{flowGen, schusterbauer2025diff2flow} offers a more efficient alternative by learning deterministic mappings with fewer denoising steps. FMRec~\citep{fmrec} applies this to sequential recommendation, while FlowCF~\citep{flowcf} adapts it to collaborative filtering. However, these models remain limited by their uni-modal Gaussian flows and lack domain-aware mechanisms, restricting their capacity to model multi-modal, domain-sensitive transitions essential for recommendation across domains.

In contrast to prior generative models that employ transformers as denoisers, resulting in long inference times, our approach leverages a Gaussian Mixture Model (GMM) as the denoiser. This avoids the computational overhead of transformer-based denoising while introducing domain-aware priors that enable expressive multi-modal modeling. Our method is scalable, intent-aware, and resilient to transition dynamics, making it well-suited for MDSR.

\begin{figure*}[t!]
    \centering
    \includegraphics[width=0.93\linewidth]{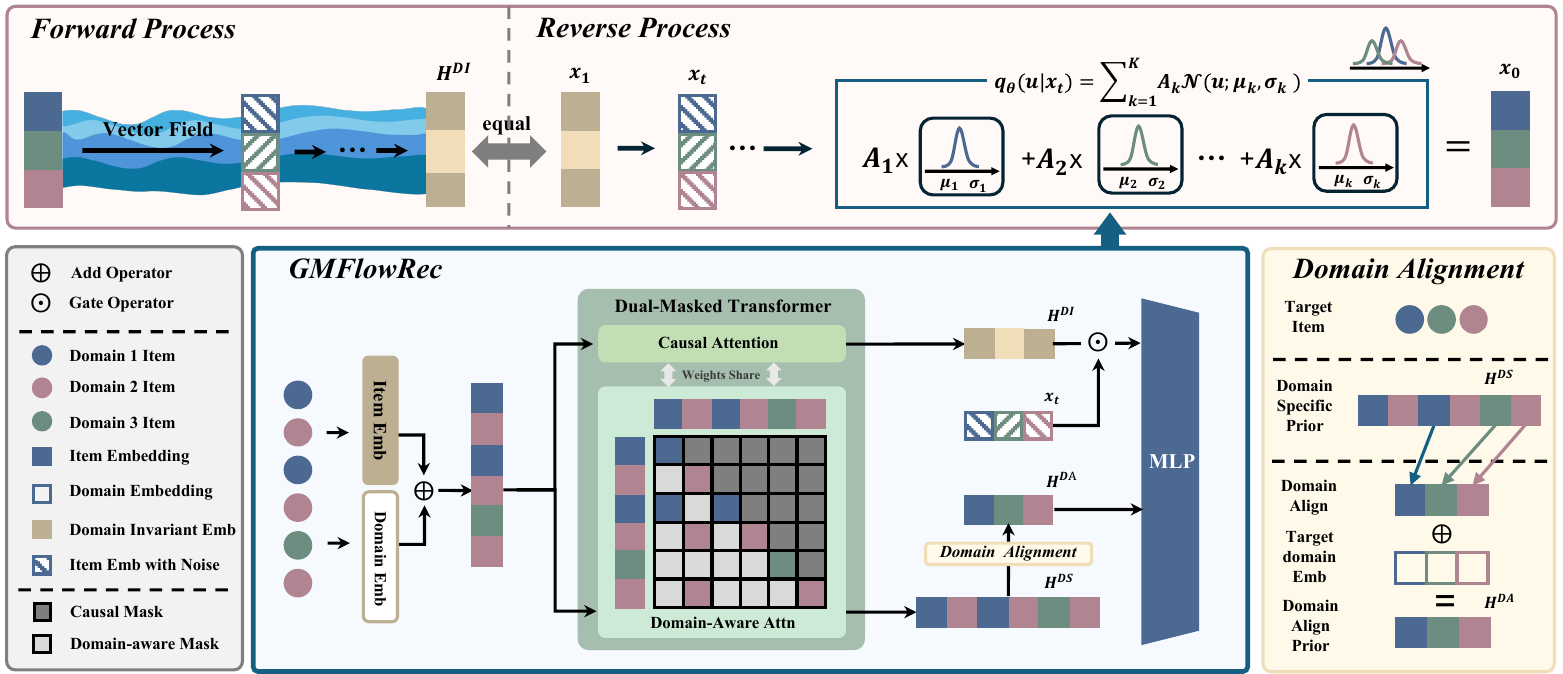}
    \caption{\textbf{GMFlowRec} integrates a \textbf{Dual-Masked Transformer} to encode multi-domain user sequences into domain-invariant and domain-specific priors. This is achieved via two complementary attention masks: a domain-invariant mask that captures global user preferences across domains, and a domain-specific mask that preserves intra-domain intent. A \textbf{Domain Alignment} operation further refines the domain-specific prior into a domain-aligned prior, which guides the reverse generative process modeled by Gaussian Mixture Flow Matching.}
    \label{fig:framework}
    \vspace{-1em}
\end{figure*}

\section{Preliminary}
\subsection{Problem Formulation}

We present the formulation of the Multi-Domain Sequential Recommendation (MDSR) task, where a user interacts with items in multiple distinct domains denoted by the set $\mathcal{D} = \{d_1, d_2, \dots, d_D\}$, with $D \geq 3$ representing the number of domains. This contrasts with Cross-Domain Sequential Recommendation (CDSR), which typically focuses on transferring knowledge between a source and target domain (often two domains) to enhance recommendations.

In MDSR, the goal is to model complex user preferences across all domains simultaneously, leveraging shared and domain-specific patterns to improve the accuracy of the recommendation. For each domain $d_k \in \mathcal{D}$, the domain-specific interaction sequence $S_{d_k}$ is defined as: $S_{d_k} = [i^{d_k}_1, i^{d_k}_2, \dots, i^{d_k}_{M}]$, $i^{d_k}_m \in \mathcal{V}_{d_k}$ where $\mathcal{V}_{d_k}$ denotes the item vocabulary specific to domain $d_k$ and $M$ is the number of interactions within a sequence. The full multi-domain sequence $S$ is constructed by aggregating all domain-specific sequences and sorting by timestamp, $S=[i_1,i_2,\dots,i_M],\;
i_m \in \mathcal{V}=\textstyle\bigcup_{d_k\in \mathcal{D}} \mathcal{V}_{d_k}$ .

For each domain $d_k$, the objective of MDSR is to predict the next item $i^* \in \mathcal{V}_{d_k}$ by modeling the joint conditional distribution over all items within domain $d_k$: $ i^* = \arg\max_{i^*  \in \mathcal{V}_{d_k}} P(i^* \mid S) \label{eq:mdsr_obj}$.

\subsection{Conditional Flow Matching}
Flow Matching (FM) \cite{flowGen}, a generative modeling framework, learns to transform a prior distribution $p_0$ into a target data distribution $p_1$ by training a velocity field $v_t$. This field defines a continuous-time trajectory $\phi_t$, where $\phi_0(x) \sim p_0(x)$ and $\phi_1(x) \sim p_1(x)$, governed by an ordinary differential equation (ODE): $\frac{d\phi_t(x)}{dt} = v_t(\phi_t(x))$. 

Conditional Flow Matching (CFM) extends this by defining a straight-line path between a source point $x_0$ and a target point $x_1$, where $x_t = (1-t)x_0 + t x_1$ for $t \in [0,1]$. The corresponding constant-velocity vector field is $u_t(x_t \mid x_1) = x_1 - x_0$. The objective is to train a neural network $v_\theta$ to approximate this field by minimizing $\mathcal{L}_{\text{CFM}}(\theta) = \mathbb{E}_{t \sim \mathcal{U}(0,1), p(x_0, x_1)} \left[ \| v_\theta(x_t, t, c) - (x_1 - x_0) \|^2 \right]$, where $c$ represents any conditioning information. In inference, starting from a source point $x_0$, one can generate a sample from the target distribution by solving the ODE from $t=0$ to $t=1$ using the learned velocity field $v_\theta$.


\section{Methodology}

\ModelName{} is a generative framework for MDSR, designed to model user preference evolution as smooth and continuous trajectories in a latent space. Existing MDSR approaches often employ separate domain-specific or multiple domain-invariant encoders, which lack transition-aware mechanisms and suffer from both scalability and coherence issues. 

In contrast, \ModelName{} introduces a unified generative framework that jointly models shared and domain-specific user intents within a single backbone Transformer, enhanced by multi-domain transition-aware trajectory generation. (i) A dual-masked Transformer disentangles domain-invariant signals, which capture global behavioral patterns, from domain-specific signals that reflect intra-domain preferences. (ii) \ModelName{} incorporates a domain-aligned prior that aligns latent intent representations with the target domain content, serving as informative priors for guiding trajectory generation. (iii) With these priors, the Multi-Domain Gaussian Mixture Flow Matching (GMFlowRec) module generates intent-aware trajectories that flexibly capture user transitions across domains. 

This generative process encodes uncertainty and variation in multi-domain behaviors, including domain shifts and new domain entries. GMFlowRec avoids the heavy parameter cost of multi-encoder MDSR designs while achieving fine-grained, transition-aware, and scalable user intent modeling. 

\subsection{Dual-masked Transformer} \label{subsec:Transformer}
MDSR requires both cross-domain generalization and fine-grained intra-domain modeling.  
For example, a user may exhibit a global preference for active lifestyles (e.g., protein supplements, running shoes, and sportswear) while maintaining distinct behavioral patterns within each domain.  Capturing this duality is critical for coherent transitions and personalized recommendations.

To address this, we construct informative priors using a dual-masked Transformer that disentangles domain-invariant and domain-specific priors, which guides a domain-aware trajectory generation.

\textbf{Dual-Masked Encoder.}  Given a multi-domain interaction sequence $\mathcal{S} = [i_1, \dots, i_M]$, each item $i_m$ is represented as $x_m = \text{Emb}(i_m) + D(d_m) + \text{Pos}(m)$. where $\text{Emb}(\cdot)$, $D(\cdot)$, and $\text{Pos}(\cdot)$ denote item, domain, and positional embeddings, respectively. The resulting sequence $X = [x_1, \dots, x_M]$ is processed by a shared Transformer encoder with two complementary attention masks:

\begin{itemize}[leftmargin=*]
\item \textbf{Domain-Invariant Prior} $H^{\text{DI}}$: Uses a standard causal mask $M^{\text{DI}}$, allowing each item to attend to all preceding items across domains, capturing the user's domain-invariant interests.
\begin{equation}\label{eq:DIPrior}
H^{\text{DI}} = \text{Encoder}(X, M^{\text{DI}}),\, \text{where} \: M^{\text{DI}}_{m,n} = \mathbb{I}[n \leq m],
\end{equation}
\item \textbf{Domain-Specific Prior} $H^{\text{DS}}$: Uses a domain-aware causal mask $M^{\text{DS}}$, restricting attention to preceding items within the same domain, preserving coherent intra-domain preferences.
\begin{equation}\label{eq:PriorDS}
H^{\text{DS}} = \text{Encoder}(X, M^{\text{DS}}),\, \text{where} \: M^{\text{DS}}_{m,n} = \mathbb{I}[d_m = d_n \land n \leq m],
\end{equation}
\end{itemize}

This efficient design disentangles global and local preferences within a single encoder, eliminating redundant per-domain networks and scaling efficiently with the number of domains.

\subsection{Domain-Aligned Prior}\label{Prior} 
Due to domain transitions, domain-specific priors may be temporally aligned but semantically misaligned across domains. For example, a user might purchase protein supplements (health) followed by running shoes (sport), which belong to different domains. To guide generation toward a specific target domain $d_{k^*}$, we construct a \textit{domain-aligned prior} $h^{\text{DA}}_{d_k}$ using the most recent in-domain interaction timestamp $m^*$:
\begin{equation}
h^{\text{DA}}_{d_k} =
\begin{cases}
H^{\text{DS}}_{m^*} + \text{Emb}(d_{k^*}), & \text{if in-domain interaction exists,} \\
\text{Emb}(d_{k^*}), & \text{otherwise (cold-start case),}
\end{cases}
\label{eq:da}
\end{equation}

This fused representation combines the last in-domain user state with the target domain context, offering adaptive initialization for domain-targeted generation, particularly beneficial for handling domain transitions and domain cold-start scenarios efficiently.

\textbf{Prior Training Objective.} To enable domain-targeted generation, we optimize the domain-aligned prior through next-item prediction. Specifically, the objective encourages the prior $h^{\text{DA}}_{u, d_k}$ to predict the next item $i^*$ within the target domain $d_k$:

\begin{equation}
\label{eq:priorloss}
\ell_{d_k}^{\text{Prior}} = -\textstyle \sum_{ \mathcal{B}_{d_k}} \log \mathcal{P}(i^* \mid h^{\text{DA}}_{u, d_k}, d_k),
\end{equation}

Here, $\mathcal{P}(i^* \mid z, d_k)$ denotes the domain-specific softmax, where $z$ is the latent representation (e.g., $h^{\text{DA}}_{u, d_k}$), and $\mathcal{V}_{d_k}$ is the item vocabulary for domain $d_k$:

\begin{equation}
\mathcal{P}(i^* \mid z, d_k) =
\frac{\exp(z \cdot \text{Emb}(i^*))}
     {\sum_{j \in \mathcal{V}_{d_k}} \exp(z \cdot \text{Emb}(j))},
\label{eq:prediction_function}
\end{equation}

This formulation aligns the prior with domain semantics, enabling robust supervision across multiple domains.

\subsection{Multi-Domain GMFlowRec}
\label{subsec:vectorfield}

Multi-domain sequences exhibit inherently uncertain transitions (e.g., moving from \textit{Health} to \textit{Sport}).  To model such dynamics, we represent recommendation as a \textbf{conditional flow}, interpolating between a user’s domain-invariant intent and the target item embedding.  The velocity field governing this transition is parameterized by a Gaussian Mixture Model (GMM) with domain-align informative prior to model diverse cross-domain trajectories.

\textbf{Vector Field Construction.}  Let $x_1 = h^{\text{DI}}_M$ denote the final domain-invariant user state, and $x_0 = \text{Emb}(i^*)$ the embedding of the target item.  We sample a timestep $t \sim \mathcal{U}(0,1)$ and construct an interpolated latent state $x_t = (1-t)x_0 + t x_1$.  This linear interpolation captures a smooth geometric transition between the user’s invariant preference and the target item, providing contextual grounding for trajectory generation. 

To stabilize early-stage training and prevent embedding collapse~\cite{ADRec}, we further introduce a fused latent representation:
\begin{equation}
\bar{x}_t = \lambda x_t + (1 - \lambda) h^{\text{DI}}_M,
\label{eq:cond_state}
\end{equation}
where $ \lambda \in [0, 1] $ modulates the contribution of the interpolated state and the domain-invariant user states. The resulting fused latent $ \bar{x}_t $ serves as a time-dependent input to the reverse flow dynamics, preserving domain-agnostic semantics while enabling robust trajectory modeling during both training and inference.

\textbf{Gaussian Mixture Parameterization.} In MDSR, user preferences are shaped by both domain-invariant and domain-aware signals. While $ \bar{x}_t $ captures a temporally grounded fusion of domain-invariant interest and target item semantics, it lacks domain-specific nuance. We thus augment it with domain-aligned prior $ h^{\text{DA}}_{d_k}$  from each domain $ d_k $, and parameterize the velocity distribution $ p(\mu \mid x_t) $ using a $ K $-component Gaussian Mixture Model:
\begin{equation}
\{\alpha_k, \mu_k, \sigma_k\}_{k=1}^K = \text{MLP}([\bar{x}_t \,\|\, h^{\text{DA}}_{d_k}]),
\end{equation}

where $ A_k = \text{softmax}(\alpha_k) $ and $ \sigma_k^2 I $ denotes spherical covariance for each component. This formulation allows the model to express uncertainty over multiple plausible next actions across domains. The resulting mixture and its mean are:
\begin{align}
\label{eq:GMM}
q_\theta(\mu \mid x_t, h^{\text{DA}}_{d_k}) &= \textstyle \sum_{k=1}^{K} A_k \mathcal{N}(\mu; \mu_k, \sigma_k^2 I), \\
\mu &= \textstyle \sum_{k=1}^K A_k \mu_k,
\end{align}
We minimize the negative log-likelihood of the true source $ x_0 $, encouraging the predicted velocity to align with the reverse trajectory:
\begin{equation}
\label{eq:gmm_loss}
\ell_{d_k}^{\text{GMM}} = \mathbb{E}_{t, x_0, x_1} \left[-\log q_\theta(\mu \mid x_t, h^{\text{DA}}_{d_k}, t)\right],
\end{equation}

\textbf{Recommendation Objective.} To ensure that the learned velocity $ \mu $ contributes to accurate next-item prediction, we align it with the recommendation objective. Specifically, we reuse the prediction function defined in Eq.~\ref{eq:prediction_function} and optimize the following loss:
\begin{equation}
\ell_{d_k}^{\text{Rec}} = - \textstyle\sum_{\mathcal{B}_{d_k}} \log \mathcal{P}(i^* \mid \mu, d_k),
\end{equation}
This objective encourages the fused and domain-aware latent dynamics to generate representations that are both semantically meaningful and predictive of user behavior across domains.

\subsection{Training and Inference}

\textbf{Overall Training Objective.}
We propose a unified generative framework for multi-domain sequential recommendation (MDSR), integrating recommendation, prior modeling, and flow-based generative guidance. The model learns domain-aware user trajectories via Gaussian Mixture Flow Matching (GMFlow), guided by informative priors and structured transitions. The overall objective is:
\begin{equation}
\mathcal{L} = \textstyle \sum_{d_k \in \mathcal{D}} 
\Big(
\ell^{\text{Rec}}_{d_k} + 
\alpha \, \ell^{\text{Prior}}_{d_k} + 
\beta \, \ell^{\text{GMM}}_{d_k}
\Big).
\label{eq:overall_loss}
\end{equation}

where $\alpha$ and $\beta$ are hyperparameters controlling the contribution of each loss term. Specifically, $\ell^{\text{Rec}}_{d_k}$ aligns the predicted velocity with next-item prediction, $\ell^{\text{Prior}}_{d_k}$ regularizes domain-aligned priors derived from user history, and $\ell^{\text{GMM}}_{d_k}$ guides the reverse flow process via Gaussian mixture modeling. The detailed training procedure is described in Algorithm~\ref{alg:training} (Appendix).


\textbf{Inference with ODE Solvers.}
During inference, we generate next-item embeddings by solving a learned flow trajectory from the user’s current state to future preferences. Unlike traditional diffusion models, our framework leverages a GM-ODE-based solver for efficient and deterministic sampling~\cite{gmflow}. Given a trained model, we begin by encoding the domain-invariant prior ($H^{\text{DI}}_m$) and the domain-aligned prior ($h^{\text{DA}}_{d_k}$). We then simulate the reverse trajectory using a first-order GM-ODE solver~\cite{gmflow}, which analytically derives the predicted velocity field ($\mu$) from the predicted Gaussian mixture, enabling accurate few-step sampling. The detailed inference procedure is Algorithm~\ref{alg:inference} (Appendix).


To make domain-specific predictions, we compute relevance scores between $\mu$ and all candidate items $ i \in \mathcal{I}_{d_k} $ in domain $ d_k $ using inner product similarity. The top-$ K $ items with the highest scores are returned as recommendations. This domain-aware scoring ensures that the generative trajectory aligns with the semantic space of the target domain, enabling accurate and efficient next-item prediction.
\section{Experiments}

 To evaluate the effectiveness and generalization capability of GMFlowRec in MDSR, we conduct extensive experiments on two large-scale datasets from Amazon and JD platforms. Specifically, we aim to answer the following research questions:
 
\begin{itemize}[leftmargin=*]
    \item \textbf{RQ1:} How does the performance of \ModelName{} compare with other SOTA baselines across different MDSR scenarios?
    \item \textbf{RQ2:} What is the impact of various components of our GMFlowRec on its performance in MDSR tasks?
    \item \textbf{RQ3:} How effectively does \ModelName{} adapt to domain transitions and the increase in domains in a user sequence?
    \item \textbf{RQ4:} Can \ModelName{} handle few-shot recommendation with limited in-domain information?
    \item \textbf{RQ5:} How does \ModelName{} compare to baselines in terms of training efficiency, inference latency, and scalability with increasing domain count and sequence length?
\end{itemize}

\begin{table*}[ht]
  \centering
\caption{Performance comparison across two multi-domain scenarios. Best results are in \textbf{bold}, second-best are \underline{underlined}. \textit{Imp.} denotes the relative improvement of GMFlowRec over the strongest baseline. All results are averaged over five runs with different seeds, reported as mean $\pm$ standard deviation. Improvements are statistically significant at the 0.01 confidence level.}
  \label{tab:performace}  
  \small
\begin{tabular}{ll |ccc| cccccc|c| c}

\toprule
 & & \textbf{SASRec} & \textbf{DiffuRec} & \textbf{FMRec} & \textbf{DREAM$_m$} & \textbf{C$^2$DSR$_m$} & \textbf{Auto++} & \textbf{DSL} & \textbf{CGRec} & \textbf{SyNC} & \textbf{GMFlowRec}& \textbf{\textit{Imp(\%)}} \\
\midrule
\multicolumn{13}{c}{\textbf{JD Dataset}}  \\
  \midrule
\multirow{4}{*}{\textbf{Domain 1}} & \textbf{H@5} & 29.81 & 32.18 & 32.07 & 32.85 & 34.87 & 31.38 & 30.43 & \underline{35.67} & 29.37 & \textbf{39.70 $\pm$ 0.02} & \textbf{11.30}\% \\
 & \textbf{H@10} & 36.23 & 39.43 & 39.44 & 40.45 & 42.45 & 37.53 & 37.50 & \underline{43.12} & 35.84 & \textbf{46.72 $\pm$ 0.05} & \textbf{8.35}\% \\
 & \textbf{N@5} & 23.72 & 25.40 & 25.37 & 25.70 & 27.51 & 25.18 & 23.76 & \underline{28.37} & 23.12 & \textbf{32.12 $\pm$ 0.05} & \textbf{13.23}\% \\
 & \textbf{N@10} & 25.79 & 27.74 & 27.76 & 28.16 & 29.96 & 27.16 & 26.04 & \underline{30.78} & 25.21 & \textbf{34.39 $\pm$ 0.02} & \textbf{11.73}\% \\
\midrule
\multirow{4}{*}{\textbf{Domain 2}} & \textbf{H@5} & 25.71 & 25.70 & 25.63 & \underline{27.70} & 26.81 & 24.84 & 22.97 & 27.65 & 21.02 & \textbf{37.73 $\pm$ 0.13} & \textbf{36.20}\% \\
 & \textbf{H@10} & 31.93 & 32.15 & 31.87 & 34.69 & 33.77 & 30.51 & 29.20 & \underline{34.70} & 26.83 & \textbf{44.09 $\pm$ 0.18} & \textbf{27.05}\% \\
 & \textbf{N@5} & 20.15 & 19.76 & 19.79 & 21.09 & 20.47 & 19.58 & 17.58 & \underline{21.35} & 15.92 & \textbf{30.82 $\pm$ 0.02} & \textbf{44.37}\% \\
 & \textbf{N@10} & 22.15 & 21.85 & 21.79 & 23.34 & 22.71 & 21.40 & 19.60 & \underline{23.61} & 17.79 & \textbf{32.87 $\pm$ 0.09} & \textbf{39.24}\% \\
  \midrule
\multirow{4}{*}{\textbf{Domain 3}} & \textbf{H@5} & 32.51 & 33.24 & 33.49 & 34.78 & 34.83 & 31.47 & 30.07 & \underline{35.87} & 28.31 & \textbf{41.57 $\pm$ 0.20} & \textbf{15.90}\% \\
 & \textbf{H@10} & 37.99 & 39.05 & 40.09 & 41.68 & 41.50 & 37.15 & 36.28 & \underline{42.14} & 34.12 & \textbf{47.81 $\pm$ 0.18} & \textbf{13.46}\% \\
 & \textbf{N@5} & 26.99 & 27.11 & 27.13 & 27.99 & 28.34 & 26.17 & 24.29 & \underline{29.27} & 23.03 & \textbf{35.03 $\pm$ 0.09} & \textbf{19.69}\% \\
 & \textbf{N@10} & 28.76 & 28.97 & 29.26 & 30.21 & 30.49 & 28.00 & 26.29 & \underline{31.29} & 24.90 & \textbf{37.04 $\pm$ 0.03} & \textbf{18.38}\% \\
\midrule
\multirow{4}{*}{\textbf{Domain 4}} & \textbf{H@5} & 25.49 & 28.19 & 28.08 & 28.21 & 29.78 & 26.96 & 25.55 & \underline{31.08} & 23.41 & \textbf{35.78 $\pm$ 0.07} & \textbf{15.11}\% \\
 & \textbf{H@10} & 31.59 & 34.76 & 34.88 & 35.43 & 36.83 & 33.08 & 32.31 & \underline{38.16} & 29.32 & \textbf{42.27 $\pm$ 0.09} & \textbf{10.78}\% \\
 & \textbf{N@5} & 20.27 & 22.05 & 21.95 & 21.98 & 23.30 & 21.69 & 19.61 & \underline{24.48} & 18.45 & \textbf{29.16 $\pm$ 0.03} & \textbf{19.13}\% \\
 & \textbf{N@10} & 22.23 & 24.17 & 24.13 & 24.31 & 25.57 & 23.68 & 21.79 & \underline{26.76} & 20.35 & \textbf{31.26 $\pm$ 0.03} & \textbf{16.81}\% \\
\midrule
\multirow{4}{*}{\textbf{Domain 5}} & \textbf{H@5} & 28.69 & 29.17 & 29.13 & 31.48 & 34.07 & 31.86 & 27.04 & \underline{35.47} & 29.27 & \textbf{41.47 $\pm$ 0.09} & \textbf{16.90}\% \\
 & \textbf{H@10} & 34.67 & 35.53 & 35.87 & 38.56 & 40.93 & 37.81 & 33.83 & \underline{42.03} & 35.21 & \textbf{47.71 $\pm$ 0.03} & \textbf{13.51}\% \\
 & \textbf{N@5} & 23.20 & 23.25 & 23.22 & 24.73 & 27.68 & 26.05 & 21.00 & \underline{28.93} & 23.67 & \textbf{34.67 $\pm$ 0.05} & \textbf{19.85}\% \\
 & \textbf{N@10} & 25.13 & 25.31 & 25.39 & 27.02 & 29.90 & 27.97 & 23.19 & \underline{31.05} & 25.59 & \textbf{36.69 $\pm$ 0.03} & \textbf{18.16}\% \\
\midrule
\multicolumn{13}{c}{\textbf{Amazon Dataset}} \\
\midrule
\multirow{4}{*}{\textbf{Health}} & \textbf{H@5} & 10.30 & 15.20 & 15.19 & 15.47 & 16.39 & 13.91 & \underline{16.56} & 16.44 & 16.45 & \textbf{17.59 $\pm$ 0.05} & \textbf{6.24}\% \\
 & \textbf{H@10} & 15.59 & 22.37 & 22.19 & 22.52 & \underline{23.63} & 20.72 & 23.61 & \underline{23.63} & 23.30 & \textbf{25.00 $\pm$ 0.03} & \textbf{5.81}\% \\
 & \textbf{N@5} & 6.85 & 10.39 & 10.31 & 10.43 & 11.30 & 9.48 & 11.30 & 11.20 & \underline{11.39} & \textbf{12.14 $\pm$ 0.03} & \textbf{6.60}\% \\
 & \textbf{N@10} & 8.54 & 12.70 & 12.57 & 12.70 & \underline{13.63} & 11.68 & 13.56 & 13.51 & 13.60 & \textbf{14.53 $\pm$ 0.02} & \textbf{6.60}\% \\
\midrule
\multirow{4}{*}{\textbf{Clothing}} & \textbf{H@5} & 9.92 & 15.32 & 15.18 & 15.80 & \underline{16.97} & 14.14 & 16.18 & 16.40 & 16.92 & \textbf{18.28 $\pm$ 0.07} & \textbf{7.74}\% \\
 & \textbf{H@10} & 14.66 & 21.74 & 21.55 & 22.96 & 23.80 & 20.58 & 23.20 & 23.48 & \underline{24.13} & \textbf{25.87 $\pm$ 0.08} & \textbf{7.19}\% \\
 & \textbf{N@5} & 6.83 & 10.68 & 10.52 & 10.95 & 11.88 & 9.90 & 11.36 & 11.47 & \underline{11.95} & \textbf{12.85 $\pm$ 0.04} & \textbf{7.55}\% \\
 & \textbf{N@10} & 8.35 & 12.75 & 12.57 & 13.26 & 14.07 & 11.97 & 13.62 & 13.75 & \underline{14.27} & \textbf{15.29 $\pm$ 0.04} & \textbf{7.18}\% \\
  \midrule
\multirow{4}{*}{\textbf{Beauty}} & \textbf{H@5} & 9.61 & 15.63 & 15.49 & 15.95 & 16.45 & 13.81 & 16.12 & \underline{16.47} & 16.16 & \textbf{17.78 $\pm$ 0.07} & \textbf{7.93}\% \\
 & \textbf{H@10} & 15.14 & 22.65 & 22.38 & 23.45 & 23.39 & 20.38 & 23.55 & \underline{24.17} & 23.35 & \textbf{25.28 $\pm$ 0.08} & \textbf{4.61}\% \\
 & \textbf{N@5} & 6.36 & 10.65 & 10.50 & 10.85 & 11.39 & 9.27 & 10.97 & \underline{11.49} & 11.10 & \textbf{12.45 $\pm$ 0.06} & \textbf{8.39}\% \\
 & \textbf{N@10} & 8.13 & 12.90 & 12.72 & 13.27 & 13.62 & 11.38 & 13.36 & \underline{13.97} & 13.42 & \textbf{14.86 $\pm$ 0.07} & \textbf{6.39}\% \\
\midrule
\multirow{4}{*}{\textbf{Grocery}} & \textbf{H@5} & 9.02 & 14.22 & 14.36 & 14.45 & \underline{15.14} & 11.89 & 15.13 & 14.77 & 13.98 & \textbf{16.85 $\pm$ 0.02} & \textbf{11.29}\% \\
 & \textbf{H@10} & 13.56 & 20.39 & 20.50 & 21.21 & 21.40 & 17.33 & \underline{21.46} & 20.72 & 20.36 & \textbf{23.60 $\pm$ 0.14} & \textbf{9.99}\% \\
 & \textbf{N@5} & 6.02 & 9.67 & 9.90 & 9.86 & \underline{10.49} & 7.97 & 10.33 & 10.15 & 9.76 & \textbf{11.97 $\pm$ 0.08} & \textbf{14.13}\% \\
 & \textbf{N@10} & 7.47 & 11.66 & 11.87 & 12.02 & \underline{12.50} & 9.71 & 12.37 & 12.05 & 11.81 & \textbf{14.14 $\pm$ 0.09} & \textbf{13.12}\% \\
\midrule
\multirow{4}{*}{\textbf{Sports}} & \textbf{H@5} & 7.00 & 11.70 & 11.53 & \underline{13.50} & 13.28 & 10.05 & 10.43 & 11.66 & 11.81 & \textbf{14.87 $\pm$ 0.11} & \textbf{10.16}\% \\
 & \textbf{H@10} & 10.85 & 17.87 & 17.67 & \underline{20.24} & 19.74 & 15.15 & 16.58 & 17.15 & 18.07 & \textbf{22.17 $\pm$ 0.37} & \textbf{9.54}\% \\
 & \textbf{N@5} & 4.62 & 8.18 & 7.84 & 9.03 & \underline{9.10} & 6.69 & 7.12 & 7.71 & 8.12 & \textbf{10.16 $\pm$ 0.08} & \textbf{11.65}\% \\
 & \textbf{N@10} & 5.85 & 10.15 & 9.81 & \underline{11.20} & 11.19 & 8.33 & 9.09 & 9.48 & 10.12 & \textbf{12.51 $\pm$ 0.07} & \textbf{11.73}\% \\

\bottomrule
\end{tabular}
\end{table*}
\subsection{Experiment Setting}

\textbf{Datasets} We evaluate \ModelName{} on two large-scale multi-domain recommendation scenarios derived from Amazon and JD e-commerce platforms. Each scenario contains user interaction sequences spanning multiple domains. To ensure meaningful cross-domain learning, we retain users with at least 10 interactions and remove items with fewer than 15 interactions. Sequences are truncated to a maximum length of 50. We use the publicly available Amazon 5-core dataset\footnote{\url{https://amazon-reviews-2023.github.io/data_processing/5core.html}} and JD-ecommerce dataset \footnote{\url{https://github.com/rucliujn/JDsearch}}. 
The dataset statistics are summarized in Table~\ref{tab:dataset_stats}; details are provided in Appendix~\ref{sec:dataset}.

\textbf{Evaluation Protocol}  We adopt a leave-one-out strategy: for each user, the last interaction is used for testing and the second-to-last for validation. Following previous works~\cite{BPR}, each evaluation case includes one positive item and 999 randomly sampled negatives from the same domain. Models rank the 1,000-item list, and we report Hit Ratio (H@K), and Normalized Discounted Cumulative Gain (N@K) $K=\{5, 10\}$.  For few-shot, domain transition, and efficiency analysis, we report Group NDCG@10,  which is evaluated by averaging NDCG@10 over all domains.

\textbf{Baselines} \ModelName{} are compared against a diverse set of baselines spanning single-domain, cross-domain, and multi-domain sequential recommendation paradigms. For fair comparison, cross-domain models are adapted to the multi-domain setting.

\noindent
\textbf{SDSR (Single-Domain Sequential Recommender)}: SASRec~\cite{SASrec}: Transformer-based model capturing intra-domain sequential dependencies. DiffuRec~\cite{diff4serec}: Models sequential interactions via a diffusion process. FMRec~\cite{fmrec}: Uses flow matching to learn sequential patterns within a single domain. 

\noindent
\textbf{CDSR$_m$ (Cross-Domain Sequential Recommender)}: C$^2$DSR$_m$ ~\cite{C2DSR}: GNN-based model with contrastive learning for cross-domain transfer. DREAM$_m$~\cite{DREAM}: Learns domain-specific and invariant preferences via supervised contrastive learning and leverages focal loss to further address the class-imbalance problem.

\noindent
\textbf{MDSR (Multi-Domain Sequential Recommender)}: CGRec~\cite{CGRec}: Mitigates negative transfer via cooperative game-theoretic modeling of domain interactions. DSL~\cite{MDSRDSL}: Learns domain-specific spaces to enhance cross-domain representation and generalization. Auto++~\cite{AutoCDSR}: Revisits self-attention with Pareto-optimal design for balanced cross-domain learning. SyNC~\cite{SyNCRec}: Introduces cooperative learning between single- and cross-domain recommenders.

\textbf{Implementation Details} \ModelName{} is implemented in PyTorch \cite{paszke2019pytorch}. We use Adam optimizer \cite{adamopt} with a learning rate of $1 \times 10^{-4}$ and a batch size of 256. Embedding dimensions are set to 64. All models are trained for 100 epochs with early stopping based on validation NDCG@10. Experiments are conducted on a machine equipped with an NVIDIA H200 GPU and 141GB RAM. Due to the space limitation, more details are included in the Appendix \ref{sec:imple}.

\subsection{Performance Comparison}
\begin{figure*}[ht]
    \centering
    \includegraphics[width=\linewidth]{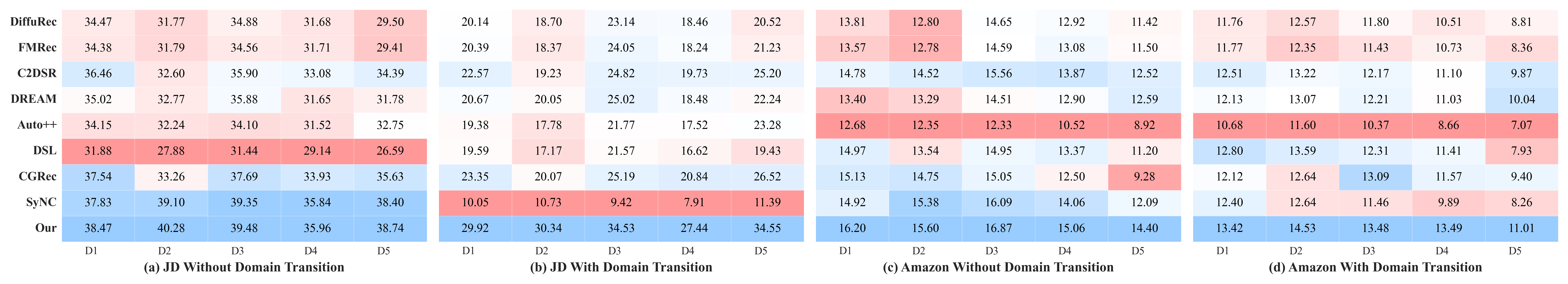}
    \caption{Performance comparison under domain transition sequence conditions across JD and Amazon datasets. We report NDCG@10 for each method under four settings: JD (w/o and w/ transition) and Amazon (w/o and w/ transition). GMFlowRec (\textbf{Our}) consistently achieves top performance, especially under transition settings. Blue shading indicates higher performance, while red indicates lower performance.}
    \vspace{-3pt} 
    \label{fig:last_diff}
\end{figure*}

\textbf{Overall Performance (RQ1)}   We evaluate GMFlowRec against three groups of baselines across multiple domains in the JD and Amazon datasets (Table~\ref{tab:performace}). The results reveal several key findings:

(i) GMFlowRec consistently achieves state-of-the-art performance across all metrics and domains. Out of 40 metric-domain combinations, GMFlowRec ranks first in every case, with statistically significant improvements under the 0.01 confidence level. The gains are especially pronounced in Domain 2 and Domain 5 of the JD dataset, where GMFlowRec surpasses the strongest baseline by up to 44.37\% and 19.85\%, respectively.

(ii) Multi-domain models outperform single-domain baselines. MDSR algorithms such as CGRec and SyNC generally outperform single-domain methods like SASRec and DiffuRec, highlighting the importance of modeling cross-domain dynamics. Interestingly, cross-domain adapted models like DREAM$_m$ and $C^2$DSR$_m$ achieve competitive performance with MDSR methods, despite being originally designed for CDSR. Their use of global and local domain modules enables effective knowledge transfer across domains. In contrast, Auto++ performs poorly in this setting, likely due to its reliance on attention regularization, which limits its ability to capture cross-domain transitions. Similarly, DSL performs reasonably well on the Amazon dataset but struggles on JD, likely due to its over-reliance on domain adaptation. 

(iii) Generative models show strong performance. Although designed for domain-isolated modeling, generative models such as DiffuRec and FMRec demonstrate robust performance compared to SASRec, highlighting the effectiveness of structured generative modeling. However, without domain-aware guidance, these models struggle to generalize across domains, unlike GMFlowRec, which explicitly captures multi-domain transitions and semantics.

\begin{table}
\small
\setlength{\tabcolsep}{1.2pt}
\centering
\caption{NDCG@10 performance comparison across two datasets.}
\label{tab:ablation}
\begin{tabular}{l|ccccc|ccccc}
\toprule
 & \multicolumn{5}{c|}{\textbf{JD}} & \multicolumn{5}{c}{\textbf{Amazon}} \\
\cmidrule(lr){2-6} \cmidrule(lr){7-11}
 & \textbf{D1} & \textbf{D2} & \textbf{D3} & \textbf{D4} & \textbf{D5} & \textbf{D1} & \textbf{D2} & \textbf{D3} & \textbf{D4} & \textbf{D5} \\
\midrule
Flow & 27.76 & 21.79 & 29.26 & 24.13 & 25.39 & 12.57 & 12.57 & 12.72 & 11.87 & 9.81 \\
GMFlow & 28.19 & 21.93 & 29.30 & 24.27 & 28.41 & 13.00 & 13.36 & 12.99 & 11.88 & 10.14 \\
v1& 32.22& 30.20& 34.74& 29.17& 33.73& 13.15& 13.32& 13.42& 12.31&10.34\\
v2& 32.47 & 31.05 & 34.81 & 29.80 & 34.15 & 13.81 & 14.03 & 13.63 & 13.01 & 11.29 \\
\midrule
\textbf{Our} & \textbf{34.57} & \textbf{32.76} & \textbf{37.21} & \textbf{31.35} & \textbf{36.67} & \textbf{14.46} & \textbf{14.91} & \textbf{14.56} & \textbf{13.98} & \textbf{12.45} \\
\bottomrule
\end{tabular}
\end{table}

\begin{figure}
    \centering
    \includegraphics[width=\linewidth]{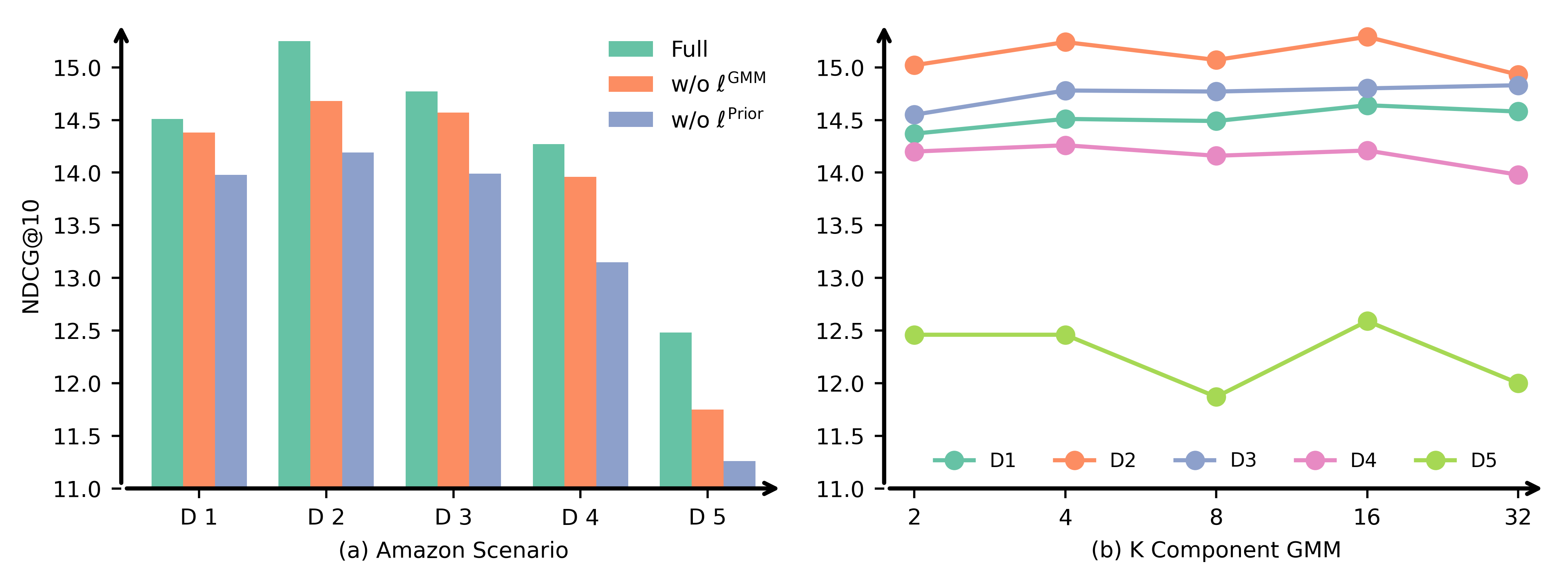} 
    \caption{
        Ablation and Component Analysis: (a) Performance comparison under the Amazon scenario, showing the impact of removing $\ell^{\text{GMM}}$ and $\ell^{\text{Prior}}$ from the full model. (b) Effect of varying the number of Gaussian components $K$ in the GMM prior.}
          \vspace{-3pt} 
    \label{fig:kgmm}
\end{figure}

\textbf{Ablation Study (RQ2)}  To assess the contribution of each component in \ModelName{}, we perform an ablation study across five domains in the JD and Amazon datasets (Table~\ref{tab:ablation}). Starting from the base Flow model, we incrementally add GMFlow, domain-aligned prior $H^{\text{DA}}$ (v1: w/o $M^{\text{DS}}$, v2: full), and the prior loss $\ell^{\text{Prior}}$, using $K=4$ Gaussian components. Each addition consistently improves performance, demonstrating the effectiveness of multi-modal modeling, domain-aware alignment, domain-specific, and prior alignment. Figure~\ref{fig:kgmm} further confirms the importance of $\ell^{\text{GMM}}$ and $\ell^{\text{Prior}}$, and shows stable performance across different values of $K$.

To further understand the contribution of each loss in our model, we conduct an ablation study under the Amazon scenario. Figure~\ref{fig:kgmm}(a) shows that removing either component leads to a consistent drop in performance across all domain groups (D1–D5). Interestingly, removing the prior loss $\ell^{\text{Prior}}$ results in a larger degradation than $\ell^{\text{GMM}}$, highlighting its critical role in aligning the generative process with domain-aware user history.

Figure~\ref{fig:kgmm}(b) shows the performance of our model with varying numbers of Gaussian components $K$ in the GMM prior. We observe that the model achieves strong results even with a small number of components (e.g., $K=4$), and performance stabilizes after. This indicates that our model is both efficient and robust, able to capture multi-modal user intents without requiring a large $K$, and not sensitive to the choice of this hyperparameter.

\begin{figure*}
    \centering
    \includegraphics[width=\linewidth]{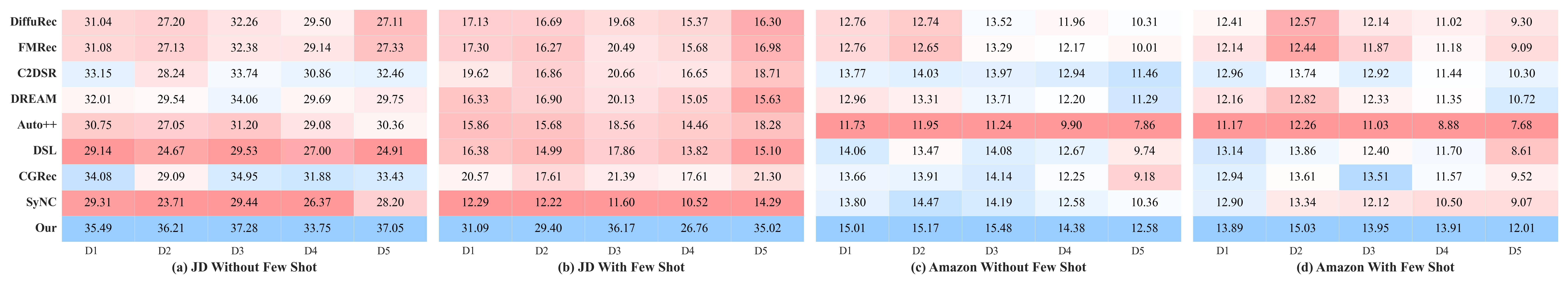}
    \caption{Performance comparison under few-shot domain sequence conditions across JD and Amazon datasets. We report NDCG@10 for each method under four settings: JD (w/o and w/ few-shot) and Amazon (w/o and w/ few-shot). GMFlowRec (\textbf{Our}) consistently achieves top performance, especially in few-shot scenarios. Blue shading indicates higher performance, while red indicates lower performance.}
      \vspace{-3pt} 
    \label{fig:few_shot}
\end{figure*}

\begin{figure}
    \centering
    \includegraphics[width=\linewidth]{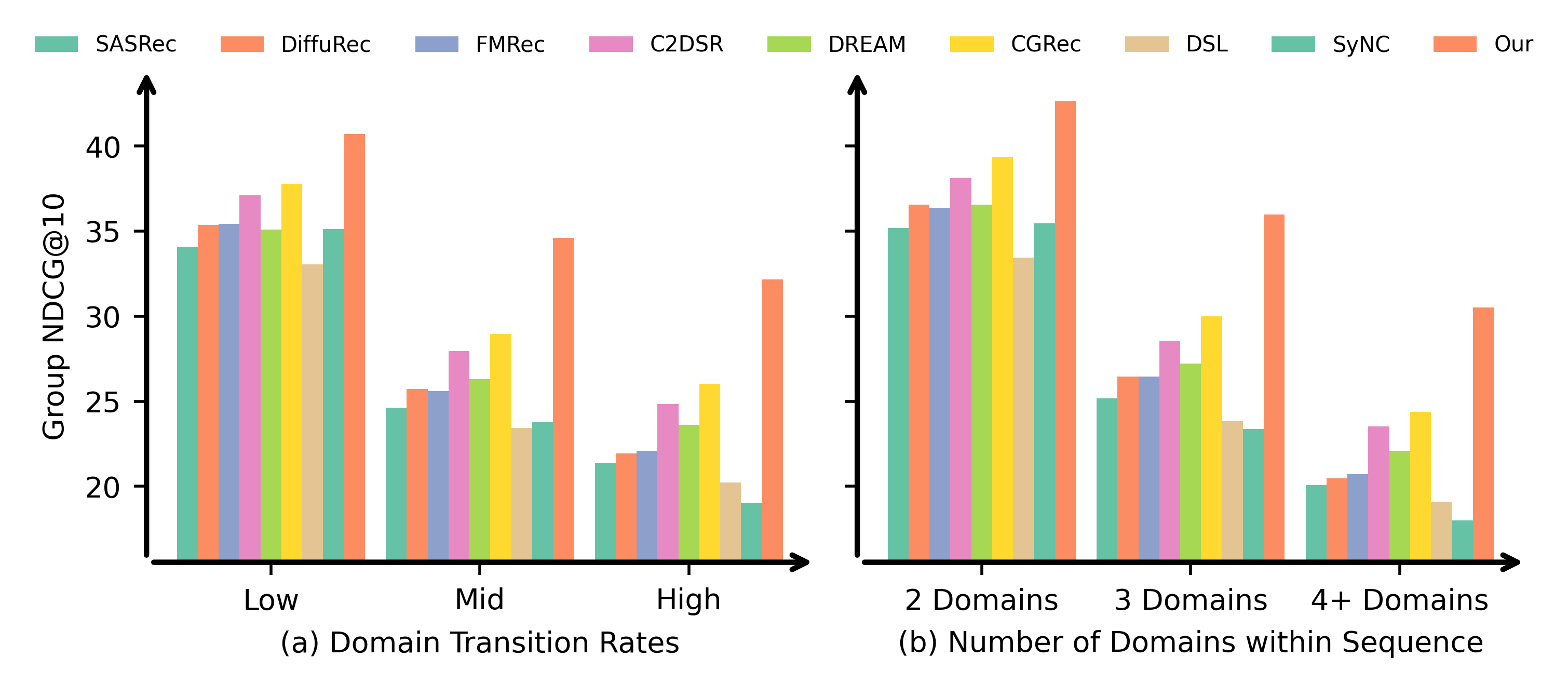}
    \caption{Performance comparison under varying domain transition rates and sequence domain counts in the JD dataset. Left (a): Group NDCG@10 across low, mid, and high domain transition rates. Right (b): Group NDCG@10 for sequences containing 2, 3, and 4+ domains. GMFlowRec consistently outperforms baselines under both high-transition and multi-domain sequence conditions, demonstrating its robustness in capturing complex cross-domain behaviors.}
      \vspace{-3pt} 
\end{figure}

\textbf{RQ3: Domain Transition Adaptation.} We investigate GMFlowRec’s ability to adapt to complex domain transitions in multi-domain sequential recommendation. In realistic scenarios, users frequently switch between domains, resulting in sparse, heterogeneous, and semantically diverse behavioral patterns. To evaluate robustness under such conditions, we design three experiments simulating increasing transition complexity:

\textit{Target Domain Transition.} As illustrated in Figure~\ref{fig:last_diff}, we examine sequences where the target domain differs from the domain of the last item. This setting introduces semantic discontinuity, making prediction more challenging. GMFlowRec demonstrates strong generalization across domains, but SyNC suffers a notable drop when the target domain changes, indicating over-reliance on the last item and vulnerability to domain transitions.

\begin{figure}
  \centering
\includegraphics[width=\linewidth]{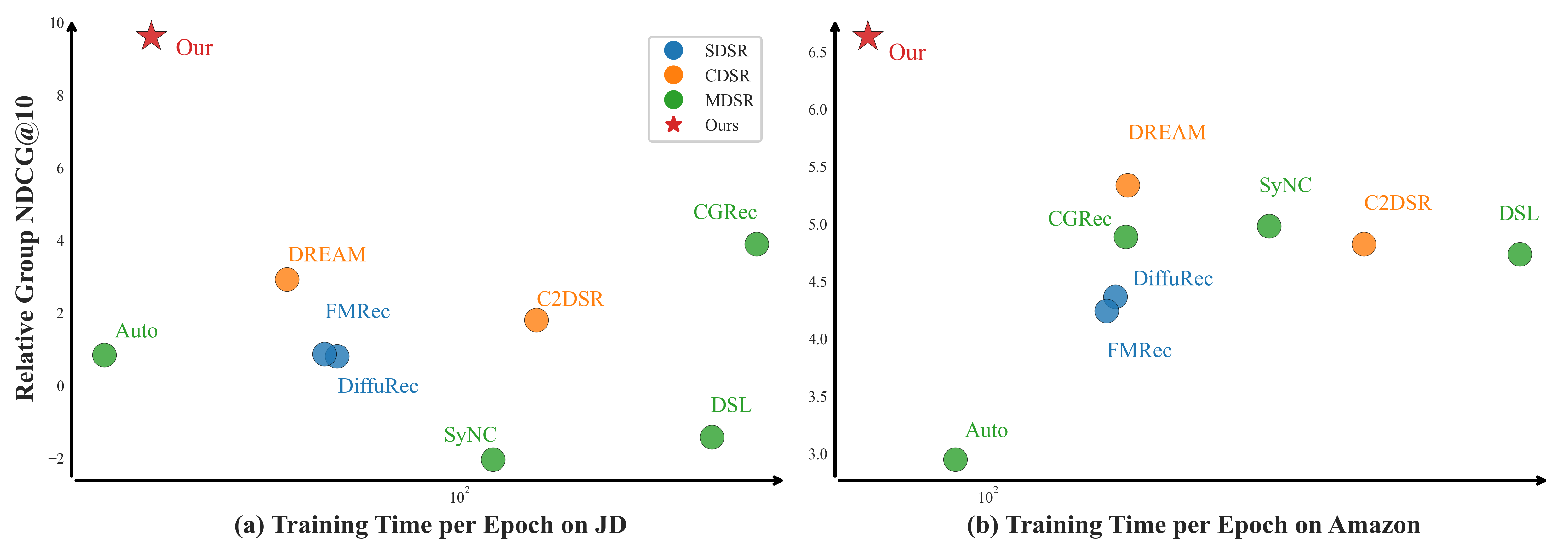} 
  \caption{Efficiency vs. effectiveness comparison on JD and Amazon datasets. Each point represents a model’s training time per epoch (x-axis) and relative group NDCG@10 (y-axis), compared with SASRec. GMFlowRec achieves superior accuracy with competitive efficiency, outperforming baselines across both datasets.}
  \label{fig:efficiency}
  \vspace{-3pt} 
\end{figure}

\textit{Domain Transition Rate.} Figure~\ref{fig:efficiency}(a) presents results under varying transition rates within sequences. Higher transition rates increase semantic heterogeneity and reduce intra-domain coherence. GMFlowRec maintains stable performance by leveraging transition-aware flow modeling, while aligned with SyNC, DSL also demonstrates a massive degradation of performance under both middle and high domain transition rates.

\textit{Number of Domains per Sequence.} We group sequences by the number of distinct domains they contain (e.g., 2, 3, or 4+ domains). As shown in Figure~\ref{fig:efficiency} (b), performance typically drops with increasing domain diversity due to sparsity and cold start effects. GMFlowRec consistently achieves top performance across all groups, confirming its scalability and resilience to domain complexity. Again, SyNC and DSL show a noticeable performance decline, reinforcing their sensitivity to domain heterogeneity.

Across all three settings on both JD and Amazon datasets, GMFlowRec consistently outperforms strong baselines. GMFlowRec maintains high recommendation quality under frequent domain shifts and an increasing number of domains, highlighting the effectiveness of its domain-aware priors and flow-based trajectory modeling.  These results validate GMFlowRec’s robustness in capturing complex cross-domain user behaviors in MDSR.

\textbf{RQ4: Few-Shot Recommendation}  We evaluate GMFlowRec’s capability in few-shot recommendation scenarios, where user interactions within the target domain are extremely limited. This setting mirrors real-world cold-start challenges, particularly in sparse or emerging domains. To systematically assess robustness under data scarcity, we conduct few-shot domain prediction experiments on the JD and Amazon datasets. Specifically, we focus on sequences with fewer than five historical interactions in the target domain.

As illustrated in Figure~\ref{fig:few_shot}, GMFlowRec consistently outperforms strong baselines across both datasets and domains. Its ability to preserve high recommendation quality under extreme data sparsity highlights the effectiveness of its domain-aligned priors and generative flow modeling. Notably, baseline models exhibit significantly degraded performance in the JD dataset under few-shot conditions, further emphasizing GMFlowRec’s advantage. These results validate GMFlowRec’s robustness in addressing cold-start and sparse-domain challenges within the MDSR framework.

\textbf{RQ5: Complexity Analysis.} To evaluate the computational efficiency of GMFlowRec, we compare its training time per epoch against baseline models on the JD and Amazon datasets. As shown in Figure~\ref{fig:efficiency}, GMFlowRec consistently achieves superior performance in terms of Relative Group NDCG@10 compared to SASRec, while maintaining competitive training efficiency. Across both datasets, GMFlowRec ranks among the top-performing models in recommendation quality, yet its training time per epoch remains lower than most baselines with inferior accuracy. This demonstrates that GMFlowRec’s architecture is not only effective but also scalable for real-world deployment. These results confirm that GMFlowRec delivers high-quality recommendations in an efficient manner, making it particularly suitable for multi-domain sequential recommendation tasks, especially with a growing number of domains.

We analyze GMFlowRec’s computational complexity with respect to the number of domains, denoted as $D$. Many MDSR models, such as DSL and CGRec, as well as CDSR-adapted variants, scale with the number of domains, resulting in a complexity of $\mathcal{O}(D+1)$, where $D$ domain-specific transformers are used alongside one shared transformer. DSL introduces an additional $\mathcal{O}(D^2)$ adapter mechanism, and CGRec applies $\mathcal{O}(D^2)$ game-theoretic comparisons, both of which are computationally intensive and difficult to parallelize. Although Auto++ uses a single transformer, it introduces information bottlenecks (IB), leading to a complexity of $\mathcal{O}((T + \text{IB})^2)$, where the number of IBs increases with $D$.

In contrast, GMFlowRec employs a \textit{single unified transformer} that is independent of domain count, enabling efficient and scalable training and inference across multiple domains. Detailed comparisons are provided in Appendix~\ref{appendix:complexity}.

\section{CONCLUSION}
In this paper, we introduced \textbf{GMFlowRec}, a generative framework for Multi-Domain Sequential Recommendation (MDSR) that combines domain-aware informative priors with Gaussian Mixture Flow Matching. GMFlowRec models user preference evolution as continuous trajectories in latent space, effectively capturing domain-specific and domain-invariant intents within a unified Transformer backbone. Extensive experiments on large-scale datasets demonstrate that GMFlowRec achieves state-of-the-art performance across diverse domains in an efficient and scalable manner. 

\bibliographystyle{ACM-Reference-Format}
\bibliography{ref}

\appendix
\newpage

\section{Dataset Description}\label{sec:dataset}

We evaluate \textbf{GMFlowRec} on two large-scale, multi-domain recommendation scenarios constructed from the \href{https://github.com/rucliujn/JDsearch}{JD-ecommerce dataset}\footnote{\url{https://github.com/rucliujn/JDsearch}} and the \href{https://amazon-reviews-2023.github.io/data_processing/5core.html}{Amazon e-commerce dataset}\footnote{\url{https://amazon-reviews-2023.github.io/data_processing/5core.html}}. For the Amazon dataset, we select five semantically related categories to simulate realistic cross-domain settings. In the JD dataset, where category names are anonymized via hashing, we choose the top five most popular categories with non-overlapping product sets.

To ensure meaningful cross-domain learning, we apply several preprocessing steps. First, we retain users with at least 10 interactions to ensure sufficient behavioral signals and filter out items with fewer than 15 interactions to reduce sparsity. Second, we sort all interactions chronologically to construct user-specific sequences. To improve computational efficiency and focus on recent user preferences, each sequence is truncated to a maximum of 50 interactions.

\begin{table}[ht]
\centering
\caption{Statistics of the two multi-domain scenarios}
\label{tab:dataset_stats}
\setlength{\tabcolsep}{3 pt}

\begin{tabular}{lrlrrr}
\toprule
 & \textbf{\#Users} & \textbf{Domain} & \textbf{\#Items} & \textbf{\#Interact} & \textbf{Sparsity} \\
\midrule
\multirow{5}{*}{\textbf{JD}} & \multirow{5}{*}{95,257} 
  & Domain 1 & 25,846 & 1,601,465 & 99.93\% \\
& & Domain 2 & 8,331 & 669,209 & 99.92\% \\
& & Domain 3 & 14,709 & 810,006 & 99.94\% \\
& & Domain 4 & 19,717 & 1,325,183 & 99.93\% \\
& & Domain 5 & 29,667 & 2,129,416 & 99.92\% \\
\midrule
\multirow{5}{*}{\textbf{Amazon}} & \multirow{5}{*}{580,329} & Health & 45,662 & 3,616,188 & 99.99\% \\
& & Clothing & 81,229 & 5,083,612 & 99.99\% \\
& & Beauty & 41,091 & 2,780,942 & 99.99\% \\
& & Grocery & 30,786 & 1,899,959 & 99.99\% \\
& & Sports & 20,078 & 928,236 & 99.99\% \\
\bottomrule
\end{tabular}
\end{table}
The datasets exhibit high sparsity while maintaining substantial diversity across domains. Specifically, the JD scenario comprises five domains, and similarly, the Amazon scenario spans five distinct domains: \textit{health}, \textit{clothing}, \textit{beauty}, \textit{grocery}, and \textit{sports}. Detailed statistics are summarized in Table \ref{tab:dataset_stats}, including the number of users, items, and interactions, as well as sparsity for each domain.

\section{Model Training and Inference Procedures}

Algorithm~\ref{alg:training} outlines the training procedure of GMFlowRec. For each multi-domain sequence, we construct domain-invariant and domain-specific priors using dual-masked attention, derive a domain-aligned prior, compute the fused latent representation, and calculate the expected velocity. The model is trained by minimizing a weighted sum of the recommendation loss, prior loss, and GMM-based flow matching loss.

Algorithm~\ref{alg:inference} describes the inference process using a GM-ODE solver. Starting from the domain-invariant and domain-aligned priors, we simulate the reverse trajectory using a GM-ODE solver to generate the next-item embedding, followed by domain-specific scoring.

\section{Implementation Details}\label{sec:imple}

We implement all models in PyTorch and adopt consistent training settings across baselines for fair comparison:

\textbf{Training Configuration:} The embedding dimension is set to 64. For sequential models, we apply a sliding window with a maximum sequence length of 50. For graph-based models, the number of GNN layers is selected from \{1, 2, 3\} based on validation performance. All models are trained using the Adam optimizer \cite{adamopt} with a batch size of 256, a dropout rate selected from \{0.1, 0.2\}, a learning rate of 0.001, and training steps of 100 with early stopping.

\textbf{Model-Specific Settings:}
\begin{itemize}[leftmargin=*]
    \item Single-domain sequential recommendation baselines are trained on all interactions without domain labels and evaluated on target-domain candidates. For models with public implementations, we follow their default settings \textbf{SASRec}\footnote{\url{https://github.com/kang205/SASRec}}, 
        \textbf{DiffuRec}\footnote{\url{https://github.com/WHUIR/DiffuRec}}, 
        \textbf{FMRec}\footnote{\url{https://github.com/FengLiu-1/FMRec}}, 
        \textbf{CGRec}\footnote{\url{https://github.com/cpark88/CGRec}}, 
        \textbf{SyNCRec}\footnote{\url{https://github.com/cpark88/SyNCRec}}, 
        \textbf{Auto++}\footnote{\url{https://github.com/snap-research/AutoCDSR}}.
    
    \item For \textbf{MDSR-DSL}, we tune $\lambda_1$ from \{0.0, 0.1, 0.5, 1\} and $\lambda_2$ from \{0, 0.01, 0.1, 1, 10, 50, 100\}.
    \item To adapt \textbf{C$^2$DSR}\footnote{\url{https://github.com/cjx96/C2DSR}} and \textbf{DREAM} to the MDSR setting, we expand the number of encoders: one shared encoder across domains and one encoder per domain. For DREAM, we replace the original sigmoid-based attention (designed for two domains) with a cross-attention module and substitute the focal loss with cross-entropy loss.
    \item \textbf{GMFlowRec (Ours):} we tune $\alpha$ from \{0.0, 0.1, 0.5, 1\}, $\beta$ from \{0, 0,00001, 0.0001, 0.01, 0.1, 1\}, and K from \{2,4,6,8, 16, 32\}. 
    We carefully tuned the model parameters of all the baselines on the validation set. Besides, all the models are carried out five times to report the average results based on different seeds.
\end{itemize}

\begin{algorithm}[ht]
\caption{Training Procedure for GMFlowRec.}
\label{alg:training}
\begin{algorithmic}[1]
\Require Multi-domain sequences $\mathcal{S}$, domain set $\mathcal{D}$, embedding function $\text{Emb}(\cdot)$, Transformer encoder $\text{Encoder}(\cdot)$, GMFlow model $f_\theta$, hyperparameter $\alpha$, $\beta$
\Ensure Trained parameters $\theta$
\For{each sequence $S$ in batch $\mathcal{S}$}
    \State Sample timestep $t \sim \mathcal{U}(0, 1)$
    \State Construct informative priors $H^{\text{DI}}$, $H^{\text{DS}} $  via Equation (\ref{eq:DIPrior})(\ref{eq:PriorDS})
    \State Construct domain-aligned prior $H^{\text{DA}}$ via Equation (\ref{eq:da})
    \State Sample fused latent representation $\bar{x}_t$  via  Equation (\ref{eq:cond_state})
    \State Compute expected velocity $\hat{v} = \sum_k A_k \mu_k$ via  Equation (\ref{eq:GMM})
    \State Compute  $ \ell^{\text{Rec}}_{d_k} $, $ \ell^{\text{Prior}}_{d_k} $, and $ \ell^{\text{GMM}}_{d_k} $ via Equation (\ref{eq:prediction_function})(\ref{eq:priorloss})(\ref{eq:gmm_loss})
    \State Update parameters $\theta$ using  Equation (\ref{eq:overall_loss})
\EndFor
\end{algorithmic}
\end{algorithm}
\begin{algorithm}[ht]
\caption{Inference Procedure with GM-ODE Solver.}
\label{alg:inference}
\begin{algorithmic}[1]
\Require Trained parameters $\theta$, number of steps $T$
\Ensure Predicted next-item $i^*$ in target domain $d_k$
\State Construct informative priors $H^{\text{DI}}$, $H^{\text{DS}} $  via Equation (\ref{eq:DIPrior})(\ref{eq:PriorDS})
\State Construct domain-aligned prior $h^{\text{DA}}_{d_k}$ via  Equation (\ref{eq:da})
\State Initial $x_1 = H^{DI}_m$ and step size $\Delta t = 1/T$ 
\For{$t = 1$ to $T$}
    \State Compute fused representation $\bar{x}_t$ via  Equation (\ref{eq:cond_state})
    \State Compute expected velocity $\hat{v}_t$ via  Equation (\ref{eq:GMM})
    \State Update latent state: $\hat{x}_{t-1} \gets \hat{x}_t + \Delta t \cdot \hat{v}_t$
\EndFor
\State Return $\hat{x}_0$
\end{algorithmic}
\end{algorithm}

\section{Computational Complexity Analysis}
\label{appendix:complexity}

\subsection{Training Complexity Analysis}
We compare the training complexity of multi-domain sequential recommendation (MDSR) and cross-domain sequential recommendation (CDSR) models by decomposing them into three components: pre-encoder, encoder, and post-encoder. For simplicity and consistency across models, we omit the embedding dimension $d$ in the encoder complexity expressions, focusing on the dominant structural factors such as the number of users $N$, sequence length $T$, and domain count $D$.

For models incorporating graph neural networks (GNNs), such as C$^2$DSR and DSL, we denote $m$ as the number of edges (i.e., user-item interactions), $n$ as the number of nodes (i.e., items), and $L$ as the number of GNN layers. The complexity of a GNN-based pre-encoder is approximated as $\mathcal{O}(L(md + nd^2))$.

Several MDSR models, including DSL and CGRec, scale linearly with the number of domains, using $D$ domain-specific encoders in addition to a shared encoder, resulting in encoder complexity of $\mathcal{O}((D+1) N T^2)$. DSL further introduces an adapter module with complexity $\mathcal{O}(D^2)$, which also appears in the post-encoder stage. DREAM require $D$ domain encoder, one share encoder, and cross-attention encoder, resulting $\mathcal{O}((D+2) N T^2)$.  SyNC requires K encoder and generally K is 4 as original paper suggested, and it may increase as the number of domain number increase.
Auto++ maintains a single encoder but introduces information bottlenecks (IB), leading to a complexity of $\mathcal{O}(N (T + \text{IB})^2)$, where the number of bottlenecks typically grows with $D$.

In contrast, our proposed GMFlowRec maintains a fixed encoder complexity of $\mathcal{O}(2 N T^2)$, independent of the number of domains. The only domain-aware component is the GMM-based prior in the post-encoder, which introduces a lightweight cost of $\mathcal{O}(K N)$, where $K$ is the number of Gaussian components (generally small). This design ensures scalability and efficiency in multi-domain settings.

\begin{table}
\centering
\setlength{\tabcolsep}{1.5pt}
\small
\caption{Model training complexity comparison across components.}
\label{tab:complexity}
\begin{tabular}{l c c c}
\toprule
\textbf{Model} & \textbf{Pre-Encoder} & \textbf{Encoder} & \textbf{Post-Encoder} \\
\midrule
C$^2$DSR & GCN: $\mathcal{O}(L(md + nd^2))$ & $\mathcal{O}((D+1) N T^2)$ & -- \\
CGRec & -- & $\mathcal{O}((D+1) N T^2)$ & -- \\
SyNC & -- & $\mathcal{O}(K N T^2)$ & -- \\
Auto++ & -- & $\mathcal{O}(N (T + \text{IB})^2)$ & -- \\
DREAM & -- & $\mathcal{O}((D+2) N T^2)$ & -- \\
DSL & \makecell[l]{GCN: $\mathcal{O}(L(md + nd^2))$ \\ Adapter: $\mathcal{O}(D^2)$} & $\mathcal{O}((D+1) N T^2)$ & Adapter: $\mathcal{O}(D^2)$ \\
\midrule
\textbf{GMFlowRec} & -- & $\mathcal{O}(2 N T^2)$ & GMM: $\mathcal{O}(K N)$ \\
\bottomrule
\end{tabular}
\end{table}

\subsection{Inference Complexity Analysis}
\begin{table}[ht]
\centering
\caption{Inference time per batch (in seconds) across JD and Amazon datasets.}
\label{tab:inference_time}
\begin{tabular}{l|c|c}
\toprule
\textbf{Model} & \textbf{JD (s)} & \textbf{Amazon (s)} \\
\midrule
DiffuRec      & 56.36& 57.20  \\
FMRec         & 42.71& 43.32\\
C$^2$DSR      & 1.04& 1.11\\
CGRec         & 0.74& 0.74\\
SyNCRec       & 4.35& 4.35\\
Auto++        & 1.39& 1.09\\
DREAM         & 1.52 & 1.32\\
DSL           & 5.74& 2.10\\
GMFlowRec (Ours) & 1.4& 1.35\\
\bottomrule
\end{tabular}
\end{table}

\begin{table}
\centering
\setlength{\tabcolsep}{1.5pt}
\small
\caption{Model Inference complexity comparison across components.}
\label{tab:inference_complexity}
\begin{tabular}{l c c c}
\toprule
\textbf{Model} & \textbf{Pre-Encoder} & \textbf{Encoder} & \textbf{Post-Encoder} \\
\midrule
DiffuRec & --& $\mathcal{O}(SN T^2)$& -- \\
FMRec & --& $\mathcal{O}(SN T^2)$& -- \\
\midrule
\textbf{GMFlowRec} & -- & $\mathcal{O}(2 N T^2)$ & GMM: $\mathcal{O}(SK N)$\\
\bottomrule
\end{tabular}
\end{table}

We analyze the inference complexity of representative MDSR and CDSR models by decomposing them into three components: pre-encoder, encoder, and post-encoder. Table~\ref{tab:inference_complexity} summarizes the theoretical inference complexity, while Table~\ref{tab:inference_time} reports the empirical inference time per batch on the JD and Amazon datasets.

To ensure a fair and realistic comparison, we adopt two practical optimizations commonly used in real-world deployments. First, we precomputed the negative sampling dataset to eliminate runtime variability introduced by stochastic sampling. Second, for models involving graph-based components (e.g., DSL, C$^2$DSR), we assume the graph structure is precomputed offline, as is standard in production systems. These settings allow us to isolate and evaluate the core inference efficiency of each model architecture.

Diffusion-based models such as DiffuRec and FMRec incur substantial inference overhead due to their iterative denoising process, where $S$ denotes the number of denoising steps. Each step typically involves a full Transformer pass, resulting in an encoder complexity of $\mathcal{O}(S N T^2)$. FMRec is relatively faster than DiffuRec due to its use of a deterministic Euler ODE solver.

In contrast, GMFlowRec achieves competitive inference efficiency with a fixed encoder complexity of $\mathcal{O}(2 N T^2)$, independent of the number of domains. Its post-encoder consists of a lightweight Gaussian Mixture Model (GMM) with complexity $\mathcal{O}(S K N)$, where $K$ is the number of Gaussian components and $S$ is the number of flow steps (typically small). This design avoids the need for domain-specific encoders or transformer-based denoisers, making GMFlowRec highly scalable and efficient in multi-domain settings.
\end{document}